\documentclass[english,notitlepage,twocolumn]{revtex4-2}
\usepackage[utf8]{inputenc}
\usepackage{color}
\usepackage{float}
\usepackage{amsbsy}
\usepackage{amstext}
\usepackage{amsmath}
\usepackage{amssymb}
\usepackage{graphicx}
\usepackage{esint}
\usepackage{setspace} 
\usepackage[english]{babel}

\begin{document}
\title{Tuning pair interactions in colloidal systems using random light fields}

\author{Augustin Muster}
\author{Diego Romero Abujetas}
\author{Frank Scheffold}
\author{Luis S. Froufe-Pérez}
 \email{Corresponding author: luis.froufe@unifr.ch}
\affiliation{%
Department of Physics, University of Fribourg, Chemin du Musée 3, 1700 Fribourg, Switzerland
}%

\begin{abstract}
We propose a method to tune interactions between absorptionless colloidal particle pairs. This is achieved via optimization of the spectral energy density of a homogeneous random optical field. Several standard and more exotic interaction potentials, as well as their negative counterparts, are shown to be successfully tuned. We show that the effective dimensionality of the space of potential functions that can be created by this means can reach up to several tens. 
\end{abstract}

\maketitle

Colloidal systems play a crucial role in both scientific and industrial fields \cite{russel1991colloidal,cosgrove2010colloid,mezzenga2005understanding}. Understanding and controlling the interactions that govern these systems is then essential for tuning their properties, such as stability and self-assembly capabilities. A conventional strategy to control these interactions relies, for instance, on modifying the ionic strength to the solvent, or adjusting the charge of the particles \cite{israelachvili2011intermolecular, hunter1987foundations,derjaguin1940repulsive}. Another powerful approach involves leveraging the transfer of momentum between light and matter. For instance, optical tweezers \cite{ashkin1997optical,ashkin1986observation,ashkin1970acceleration,jones2015optical} use a highly focused laser beam to precisely control the position and movement of particles. Beyond simple trapping, light can also induce additional interactions, such as optical binding, in which multiple particles interact through the scattered fields of an incident beam \cite{burns1990optical,thirunamachandran1980intermolecular,dholakia2010colloquium,zhang2024determining}.

While forces induced by deterministic light fields are generally anisotropic and not translation invariant, fluctuating electromagnetic fields can generate interactions that are both isotropic and translation invariant. A prominent example are dispersion forces such as Casimir, Casimir-Polder, and Casimir-Lifshitz interactions \cite{casimir1946influence,casimir1948influence,buhmann2013dispersionI,buhmann2013dispersionII}, which are typically attractive but may become repulsive under certain conditions \cite{kenneth2002repulsive}. Closely related, even coherently scattered thermal radiation induces weak  attractive forces between atoms and macroscopic objects \cite{Haslinger2017Attractive}. More recently, theoretical and experimental studies \cite{boyer1973retarded,brugger_controlling_2015,luis2022active} have demonstrated that artificially generated random optical fields can be engineered to produce either attractive or repulsive interactions, depending on their spectral energy density. This establishes the possibility of tuning colloidal pair interactions by tailoring the spectrum of the underlying fluctuating field.

Under this perspective, it has been recently shown that a careful design of the spectral energy density of the random field can lead to the suppression of optically induced pairwise interactions \cite{muster2025pure}, even at high total energy densities, thereby enabling the realization of purely many-body interactions.

Here, we develop and analyze a systematic method to design the spectral energy density $u_E\left(\omega\right)$ of a random electromagnetic field that induces a prescribed pair interaction potential. The paper is organized as follows, in section \ref{section_model}, we present the statistical model  of the random field and describe the induced interactions, at different frequencies, on pairs of dielectric particles presenting electric and magnetic dipole excitations. 

In section \ref{fitting_potentials}, we consider the design of the spectral energy density $u_E\left(\omega\right)$ as a constrained quadratic programming problem. We demonstrate that it can be solved efficiently using the nonnegative least squares (NNLS) algorithm. We apply this method to the design of spectral densities leading to a few representative potentials and their negative counterparts.

In section \ref{dimensionality} we generalize the electromagnetic response of the scatterers by means of a Lorentzian electric and magnetic polarizability. We study the effective dimensionality, given the non-negativity constraint, of the space of potential functions that can be created by tuning the energy density spectrum of the random field.

\section{Pair interactions induced by artificial random light fields  \label{section_model}}

We consider a  pair of particles illuminated by an artificial random light field. At each frequency $\omega$ the random field is a superposition of plane waves with random wave vectors $\mathbf{k}$ and polarization states that are homogeneously and isotropically distributed. According to \cite{setala_spatial_2003,brugger_controlling_2015} the cross-spectral tensor of such a field is proportional to the imaginary part of the  dyadic electric Green tensor, $G_E$, in the homogeneous host medium
\begin{equation}
\langle\mathbf{E}_0\left(\mathbf{r},\omega\right)\mathbf{E}_0^\dagger\left(\mathbf{r}',\omega'\right)\rangle=\frac{8\pi U_E}{\epsilon_0\epsilon_hk}\text{Im}\left\{G_E\left(\mathbf{r},\mathbf{r}'\right)\right\}\delta\left(\omega-\omega'\right),
\end{equation}
where $\epsilon_h$ is the permittivity of the medium. The average electric energy density $U_E$ is a function of the average squared amplitude $\langle|E_0|^2\rangle$ of the plane waves generating the random field $U_E=\frac{1}{2}\epsilon_0\epsilon_h\langle|E_0|^2\rangle$.

In the reminder of this paper, we shall model the electromagnetic response of each colloidal particle as induced electric and magnetic dipoles with scalar polarizabilities $\alpha_d$ ($d=e,m$). The discussion presented here could nevertheless be generalized to more complex responses.

Following \cite{brugger_controlling_2015}, the pair interaction potential $U\left(r\right)$ between the two absorptionless particles that is induced by the artificial random light field reads
\begin{equation}
    U(r)=\int_0^\infty u_E\left(\omega\right)V\left(r,\omega\right)d\omega.
    \label{master-formula}
\end{equation}
Here, $V\left(r,\omega\right)$ describes the induced interaction at a single frequency and reads
\begin{equation}
        V\left(r,\omega\right) =\frac{2\pi}{k^3}\text{ImTr}\left[\mathbb{I}-k^4G\left(\textbf{r}_1,\textbf{r}_2,\omega\right)\alpha G\left(\textbf{r}_2,\textbf{r}_1,\omega\right)\alpha\right],
        \label{V-equation}
\end{equation}
where $G$ is the $6\times6$ complex Green's tensor in the homogeneous medium \cite{muster_coupledelectricmagneticdipolesjl_2025} and alpha is the $6\times6$ diagonal complex matrix defined as $\alpha\equiv\text{diag}\left(\alpha_e,\alpha_e,\alpha_e,\alpha_m,\alpha_m,\alpha_m\right)$.

As a particular example of colloidal particles, we consider spherical silicon  particles of radius $a=230 nm$ in the infrared region ($\epsilon=12$) immersed in water $\epsilon_h=1.77$,  and separated by a center-to-center distance $r$. Light scattering in the region $\lambda \in \left [1.2 ,2.0 \right ]\mu\text{m}$ is well described \cite{garcia-etxarri_strong_2011} by induced electric and magnetic dipoles with polarizabilities $\alpha_e=6\pi i a_1/k^3$ and $\alpha_m=6\pi i b_1/k^3$, respectively, where $a_1$ and $b_1$ are the first Mie coefficients \cite{bohren_absorption_2008, hulst_light_2012}.

Let us consider an artificial random light field at a single frequency $\omega_0$, i.e. $u_E\left(\omega\right)=U_E\delta\left(\omega-\omega_0\right)$. Equation \eqref{master-formula} therefore reduces to 
\begin{equation}
    U\left(r\right)=U_E V\left(r,\omega_0\right). 
    \label{single_int}
\end{equation}

\begin{figure}
    \centering
    \includegraphics{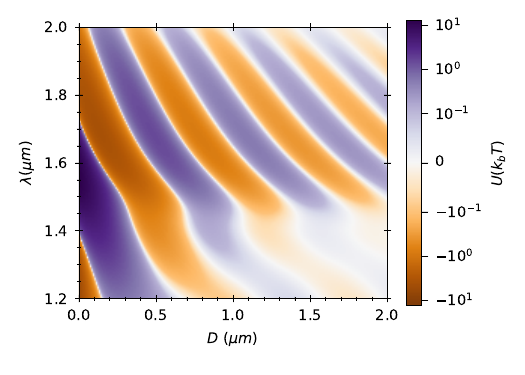}
    \caption{Pair interaction potential $U$ induced by a random field at a single frequency as a function of its wavelength $\lambda$ and the center-to-center distance $D$ of the the two dielectric particles of radius $a=230nm$ immersed in water. $T$ is 298K and the energy density of the random field is $U_E= 10^{-17} J\cdot\mu m^{-3}$.}
    \label{fig:single_pots}
\end{figure}

Figure \ref{fig:single_pots} shows the pair interaction potential between the silicon particles as a function of their surface to surface distance $D=r-2a$ and the wavelength of the monochromatic artificial random electromagnetic field. The interaction potential is an oscillating function both as a function of wavelength and distance, while showing some algebraic decay with distance. It is worth noticing that the relative position of dipole electric and magnetic resonances determines the exact character of the interaction at each point \cite{brugger_controlling_2015}, being attractive or repulsive. This, in turn, will allow for the design of energy spectra leading to prescribed potential functions.

\section{Design of arbitrary potentials \label{fitting_potentials}}
The variety of pair interaction potentials that can be obtained by monochromatic spectral energy density (Figure \ref{fig:single_pots}) suggests that these potentials can be combined in order to tune the pair interaction $U\left(r\right)$ to make it fit to an arbitrary target $U_t\left(r\right)$  by tailoring the spectral energy density. We consider a set of monochromatic artificial random light fields at frequencies $\omega_i=2\pi c/\lambda_i,i=1,..N_l$ with their associated energy density $U_E^i$. Each of these monochromatic random fields is inducing an optical interaction $U_i\left(r\right)$ given by equation \eqref{single_int}. In this section, we take $N_l=200$ different monochromatic random fields linearly spaced in terms of wavelength in an interval $\lambda_i=\left[1.2\mu m,2.0 \mu m\right]$. The resulting interaction being the linear superposition of each monochromatic component

\begin{equation}
    U\left(r\right)=\sum^{N_l}_{i=1}U_i\left(r\right)=\sum^{N_l}_{i=1}U_E^iV\left(r,\omega_i\right) \label{discrete_pot}.
\end{equation}

To fit the potential (eq.(\ref{discrete_pot})) to the target interaction $U_t\left(r\right)$, we consider a center-to-center distance interval $r\in\left[r_e+2a,D_{max}+2a\right]$ where $r_e$ is an exclusion radius below which the interaction is not considered. Introducing $r_e$ avoids using the dipole response model below distances (typically  $r<2a$) where it may start fo fail. In this case a detailed scattering model would be needed, although the main results of this paper should not be jeopardized.

We sample the potential using equally spaced $N_d$ distances $r_j$, $j=1,...,N_d$. We define a loss function to be minimized

\begin{equation}
    \tilde{\chi}^2=\sum_{j=1}^{Nd}\left(\sum^{N_l}_{i=1}U_E^iV\left(r_j,\omega_i\right)-U_t\left(r_j\right)\right)^2.
\end{equation}
This is a quadratic programming optimization problem \cite{bazaraa2006nonlinear,luenberger1984linear} where the optimization parameters are the energy densities for each frequency with the obvious non-negativity constraint $U_E^i\ge0$ . This nonnegative least squares (NNLS)  minimization problem \cite{nnls} can be solved numerically using the algorithm described in \cite{nnls_algo} and implemented in the SciPy library \cite{2020SciPy-NMeth}. In order to quantify the quality of the optimization procedure, we compute the error defined as
\begin{equation}
    \text{Error}=\frac{\sqrt{\sum_j^{N_d}\left(U\left(r_j\right)-U_t\left(r_j\right)\right)^2}}{\sqrt{\sum_j^{N_d}U_t\left(r_j\right)^2}} \label{error_def}.
\end{equation}

In the next subsections, we solve the optimization problem for a few particular potentials and discuss the feasibility of the method depending on the different parameters.

\subsection{Electrical double-layer potential}
\label{DL}
Charged colloidal particles in water are experiencing screened Coulomb interactions due to the ions always present in the liquid. This interaction is called the electrostatic double layer interaction \cite{israelachvili2011intermolecular,jones2002soft}. In the Debye-Hückel theory \cite{debye1923theorie}, i.e. when the electric potential is small, the double layer pair interaction potential reads
\begin{equation}
    U^{DL}\left(r\right)= \Phi e^{-\kappa r},
\end{equation}
where $\kappa^{-1}$ is the Debye screening length. 

While it is possible to tune $\kappa$ and $ \Phi$  by altering the ionic strength of the medium or the charge states of the colloid \cite{israelachvili2011intermolecular, hunter1987foundations,derjaguin1940repulsive}, it would be interesting to induce or cancel the double layer interaction by using interactions caused by random light fields.

To show this possibility, we apply the minimization procedure to both positive (to induce the interaction) and negative (to cancel it) versions of the double layer interaction $U^{DL}\left(r\right)$ and $-U^{DL}\left(r\right)$ respectively. We use different exclusion radii $r_e\in\left[0\mu m,0.5\mu m\right]$,  and inverse screening lengths $\kappa\in\left[0.3\mu m^{-1},30\mu m^{-1}\right]$. 

\begin{figure*}
    \centering
    \includegraphics[width=1.5\columnwidth]{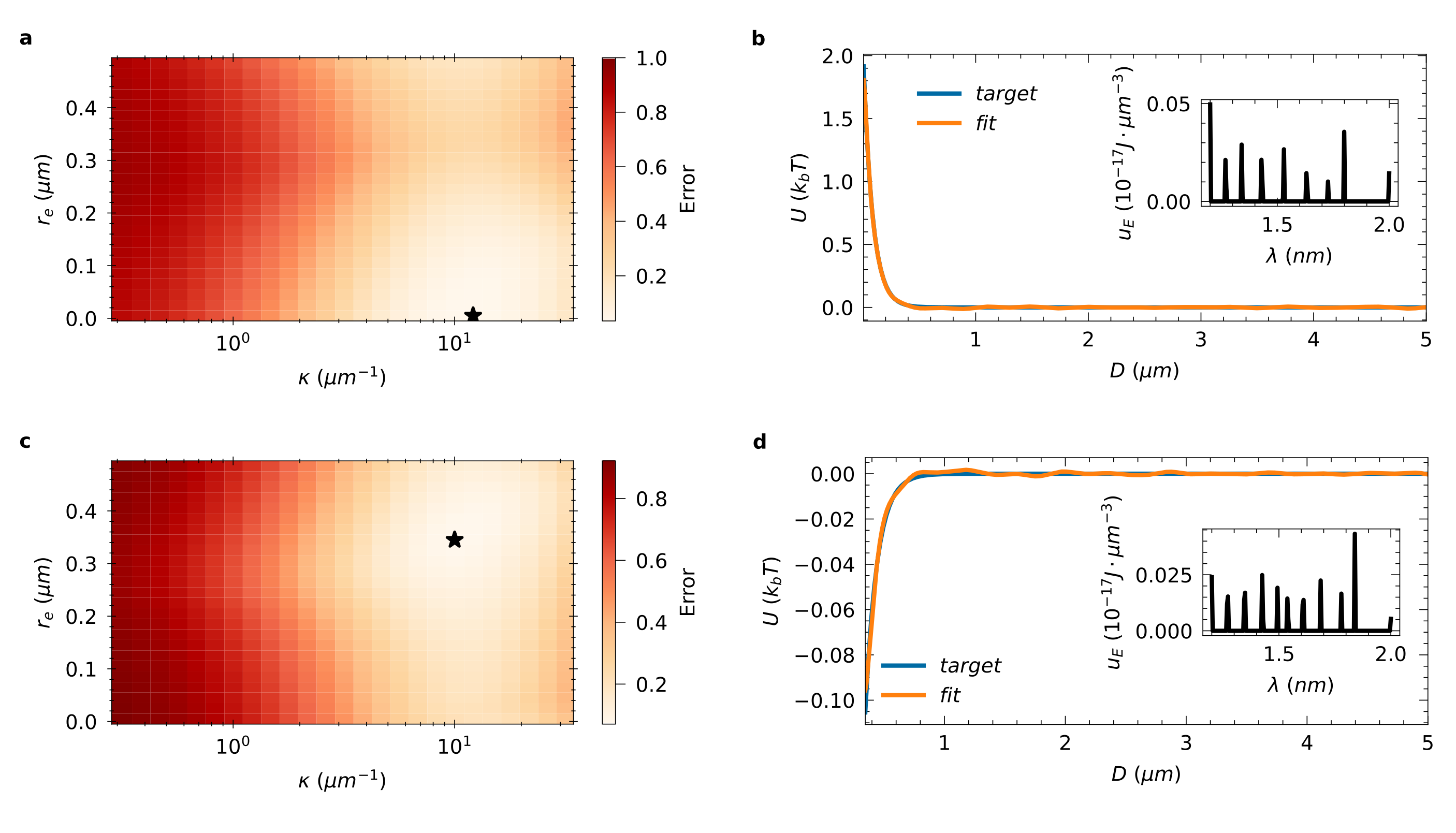}
    \caption{\textbf{a}, color map of the error of the fitting procedure for $U^{DL}\left ( D \right)$ as a function of the exclusion radius   $r_e$ and the parameter $\kappa$. \textbf{b}, comparison of the obtained and target potentials, as a function of the surface-to-surface distance $D$, for the set of parameters leading to the smallest error (star in \textbf{a}, $\kappa=12.155\mu m^{-1}$, $r_e=0\mu m$). In the inset, the obtained energy density of the random field is shown. \textbf{d} shows the error map obtained for $-U^{DL}\left ( D \right)$, correspondingly, \textbf{c} compares the target potential with the best fit (star in \textbf{c}, $\kappa=10\mu m^{-1}$, $r_e=0.370\mu m$), together with the optimized energy density spectrum. In all cases $\Phi=33k_BT$, and $T=298K$.}
    \label{fig:fig_DL}
\end{figure*}

Figure \ref{fig:fig_DL} shows the error (eq. \ref{error_def}) obtained with the optimization procedure on $U^{DL}\left(r\right)$ (\ref{fig:fig_DL}\textbf{a}), and its negative counterpart, $-U^{DL}\left(r\right)$ (\ref{fig:fig_DL}\textbf{c}). Both panels show that the double layer interaction potential can be well fitted in the range of surface-to-surface distances $D\le 5.0 \mu \text{m}$ ($D=r-2a$), and for all tested exclusion radii and inverse screening lengths greater than $2\mu \text{m}^{-1}$. Figures \ref{fig:fig_DL}\textbf{b,d} illustrate two situations where the double layer pair interaction potential can be better approximated and, respectively, canceled using interactions induced by random optical fields. As can be seen in these panels, the difference between the target potentials and the optically induced ones is well below the thermal energy $k_B T$.

The corresponding optimized energy density spectra are shown in the insets of Fig.(\ref{fig:fig_DL}\textbf{b,d}). It is remarkable that the energy densities $U_E^i$ obtained by solving the NNLS problem are, for most of them, not contributing to $U\left(r\right)$. Only about ten out of a thousand are nonzero. This effect seems to be a consequence of the Karush-Kuhn-Tucker (KKT) theorem for non-linear programming \cite{karush2013minima,kuhn2013nonlinear}. Under rather general circumstances, the minimum lies at the boundary of the constrained zone, where most of the parameters will be zero, with only a few of them being strictly positive. Hence, the spectral energy density needed to induce a prescribed interaction seems to be naturally composed of a few narrow lines. Notice that the scale of energy densities used in all computations throughout this work are in the range $U_E\in \left [ 10^{-19},10^{-14}\right ] \text{J}/ \mu \text{m}^3$, which is within reach of conventional sources \cite{,brugger_controlling_2015}. 


\subsection{Lennard-Jones potential}
Similar calculations as the ones presented in the previous subsection have been carried out using the Lennard-Jones (LJ) \cite{jones1924determination1,jones1924determination2,100yearsLJ} potential as a target interaction. The LJ potential can be written as 
\begin{equation}
    U^{LJ}\left(r\right)= \Phi \left[\left(\frac{\sigma}{r}\right)^{12}-\left(\frac{\sigma}{r}\right)^6\right],
\end{equation}
where $\sigma$ sets the length scale of the interaction.

\begin{figure*}
    \centering
    \includegraphics[width=1.5\columnwidth]{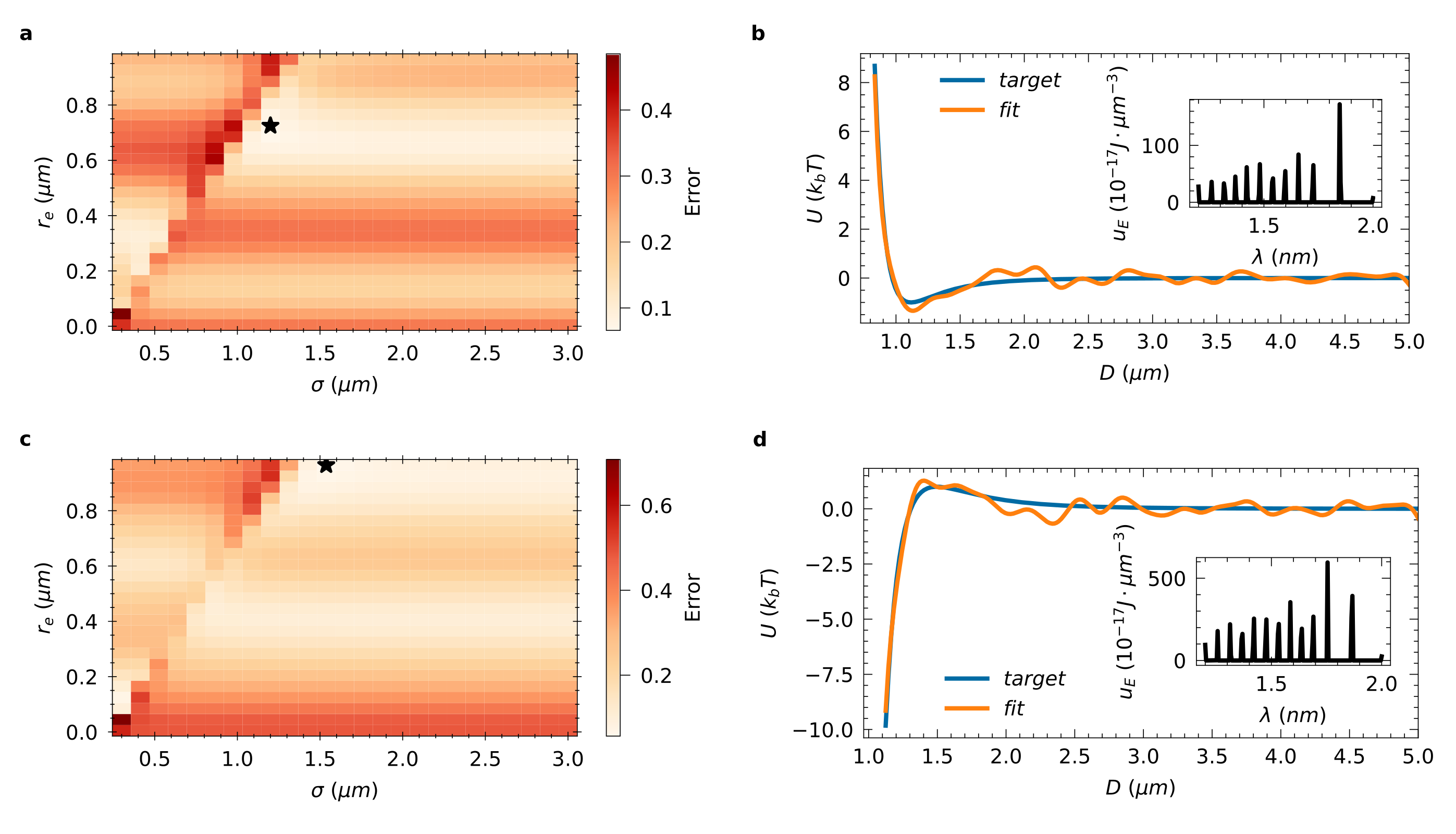}
    \caption{\textbf{a,c}, color map of the error in the fitting procedure for $U^{LJ}\left ( D \right)$ and $-U^{LJ}\left ( D \right)$ resp. as a function of the exclusion radius $r_e$ and the parameter $\sigma$. \textbf{b,d} show the best fittings (stars) in \textbf{a} ($\sigma=1.2\mu m$,  $r_e=0.72\mu m$ ) and \textbf{b} ($\sigma=1.538\mu m$,  $r_e=0.96\mu m$ ), compared with the corresponding target potentials.  We show the optimized spectral energy in the insets. In all cases $\Phi=4k_BT$, $T=298K$.}
    \label{fig:figure_LJ}
\end{figure*}

Figures \ref{fig:figure_LJ}\textbf{a,c} present the error of the fitting procedure on $U^{LJ}$, respectively $-U^{LJ}$, for a set of exclusion radii $r_e\in\left[0\mu m,1\mu m\right]$ and potential parameters $\sigma\in\left[0.3\mu m,3\mu m\right]$ as well as $\Phi=4k_BT$. It is shown that both potentials can be fairly well fitted in a relatively wide region of parameters. 
The best fits for both $U^{LJ}\left ( D \right )$ and $-U^{LJ}\left ( D \right )$ are shown in fig. \ref{fig:figure_LJ}\textbf{b,d} together with the corresponding optimized spectral energy density. While some errors appear very close to contact, the general shape of the rest of the potential is well reproduced with an error of less than $k_BT$, in particular in the zone of the potential core.

\subsection{Oscillatory potential}
As a third example, we solve the NNLS problem with a more exotic target  potential defined as
\begin{equation}
    U^{SHU}\left(r\right)=\Phi \frac{j_1\left(k_c r\right)}{k_c r},
\end{equation}
where $j_1$ is the first order spherical Bessel function of the first kind and the reciprocal length $k_c$ controls the range of the potential. This potential is interesting because it allows generating stealthy hyperuniform (SHU) points patterns in 3D \cite{TORQUATO20181,PRLhyper2016}. SHU points patterns are defined by their structure factor $S(k)$ which is constrained by $S(k<k_c)=0$ and have remarkable optical properties due to their correlated-disordered nature \cite{RevModPhys.95.045003}. However, this kind of pair potential does not occur naturally.

\begin{figure*}
    \centering
    \includegraphics[width=1.5\columnwidth]{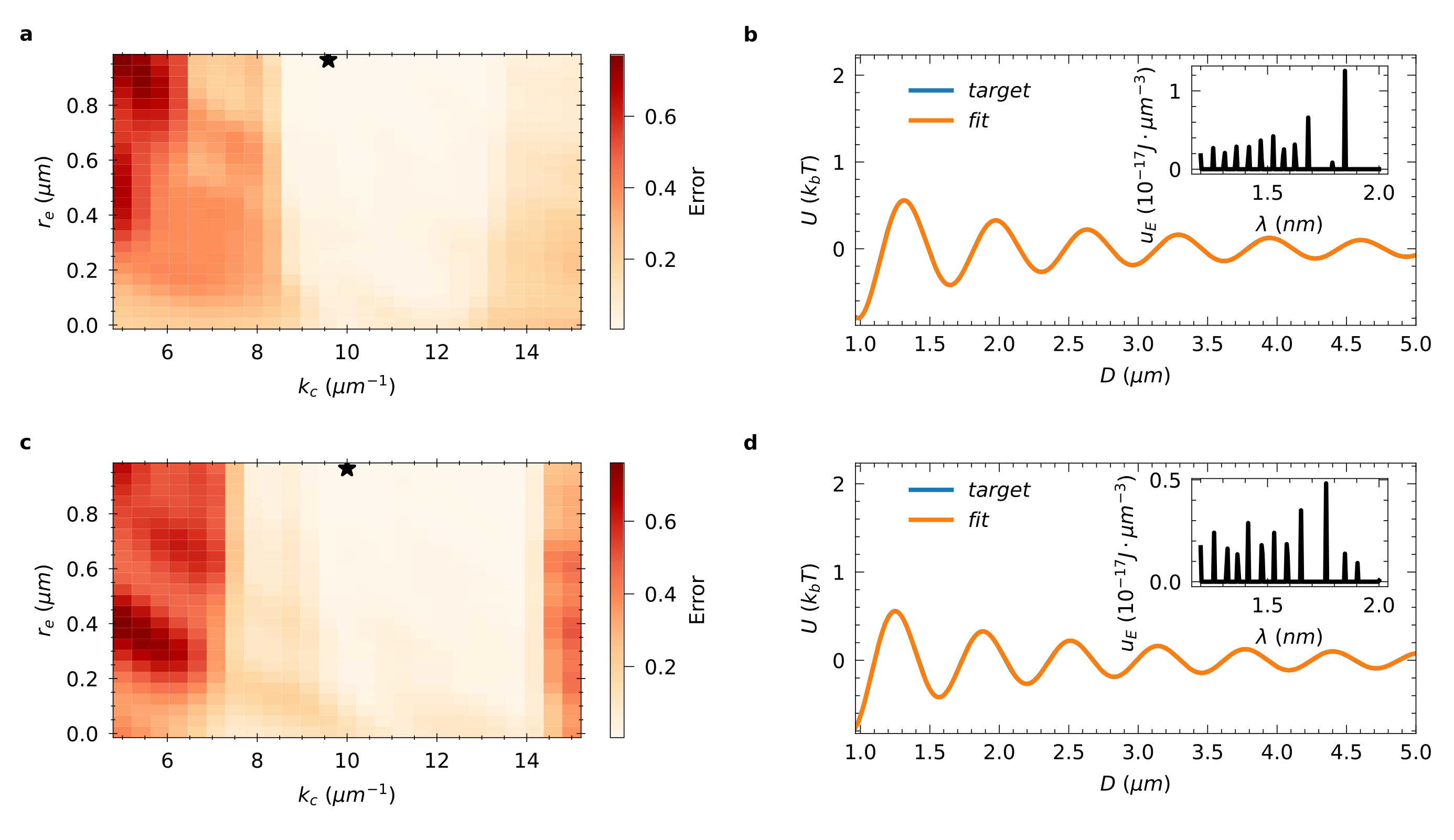}
    \caption{\textbf{a,c}, color map of the error in the fitting procedure for $U^{SHU}\left ( D \right)$ and $-U^{SHU}\left ( D \right)$ resp. as a function of the exclusion radius $r_e$ and reciprocal length $k_c$. \textbf{b,d} respectively show the best fittings (stars) in \textbf{a} ($k_c=9.58\mu m^{-1}$,  $r_e=0.96\mu m$ ) and \textbf{b} ($k_c=10 \mu m^{-1}$,  $r_e=0.96\mu m$ ), compared with the corresponding target potentials.  We show the optimized spectral energy in the insets. In all cases $\Phi=40k_BT$, $T=298K$.}
    \label{fig:figure_SHU}
\end{figure*}

We assess the possibility of inducing a $U^{SHU}\left(r\right)$ potential in this subsection. Figure \ref{fig:figure_SHU}\textbf{a,c} presents the error of the fitting procedure on $U^{SHU}$ and $-U^{SHU}$ respectively, for a set of exclusion radii $r_e\in\left[0\mu m,1\mu m\right]$ and potential parameters $k_c\in\left[5\mu m^{-1},15\mu m^{-1}\right]$. The error is fairly well minimized in the range $k_c\in \left [8 \mu\text{m}^{-1}, 14 \mu\text{m}^{-1}\right ]$, and almost independent of the exclusion radius. Figure \ref{fig:figure_SHU}\textbf{b,d} shows the comparison between the target and fitted potentials with parameters corresponding to the minimal error for $U^{SHU}$ and $-U^{SHU}$ respectively, together with the corresponding energy density spectra in the insets.

\section{Dimension of the fittable subspace of potentials \label{dimensionality}}

Giving a quantitative estimate of the actual variety of interactions that can be induced by random light fields with arbitrary spectral energy density is far from being trivial. On the one hand, the non-negativity condition on the energy density of each line $U_E^i$ renders the minimization problem non-analytical even formally. On the other hand, even removing the positivity condition, the dimensionality of the available space is not necessarily the number of base functions $N_l$ since adding many more wavelengths within the same interval will not provide more usable degrees of freedom.
Added to these limitations, we have to also consider the different possibilities in the electromagnetic response of the particles. Even in the simplified case of scatterers described by their polarizabilities $\alpha_{e,m}$, the quantity of parameters and its complex relation precludes an exact description of the achievable induced potentials. 

In this section we consider a set of possible potential functions as a vector space with a number of dimensions that we shall estimate in two different steps and for particles with a dipole response not restricted to the previously considered silicon ones. 

We consider both electric and magnetic polarizabilities to be Lorentzian with a quality factor $Q_{e,m}$. For the sake of simplicity $Q_{e}=Q_{m}=Q$, here, all wavelengths will be scaled by the electric resonance one $\lambda_0$, and there is a detuning $\Delta$ accounting for the relative difference in the resonance frequency of the magnetic excitation with respect to the electric one. This parametrization is described in more detail in \cite{muster2025pure} and summarized in Appendix B. We notice that, in this work, only absorptionless responses are considered. Hence, the polarizabilities fulfill the optical theorem.

In order to obtain an upper bound of the dimensionality of the space of achievable potential functions, we temporarily remove the non-negativity condition on the individual energy densities. By doing so, the minimization problem becomes strictly linear with a dimension given by the dimension of the vector space given by equation (\ref{discrete_pot}). In order to avoid quasi-degeneracies caused by the presence of many similar functions in the basis, we consider an effective dimensionality understood as the number of relevant singular values of the matrix $A$ with elements $A_{ij} = U_i\left(r_j\right)$, for a set of wavelengths $\lambda_i$, with $ i = 1, \dots, N_l $, and distances $ r_j $, with $ j = 1, \dots, N_d $. 

We compute the singular value decomposition  (SVD) of A, and we sort the obtained singular values in descending order. The dimension $d$ of the space of attainable functions is then determined by counting the number of singular values $\sigma _i$ whose ratio to the biggest one exceeds a given threshold \( t \), i.e. $\sigma _i/\sigma_1\ge t$.

\begin{figure*}
    \centering
    \includegraphics[width=1.5\columnwidth]{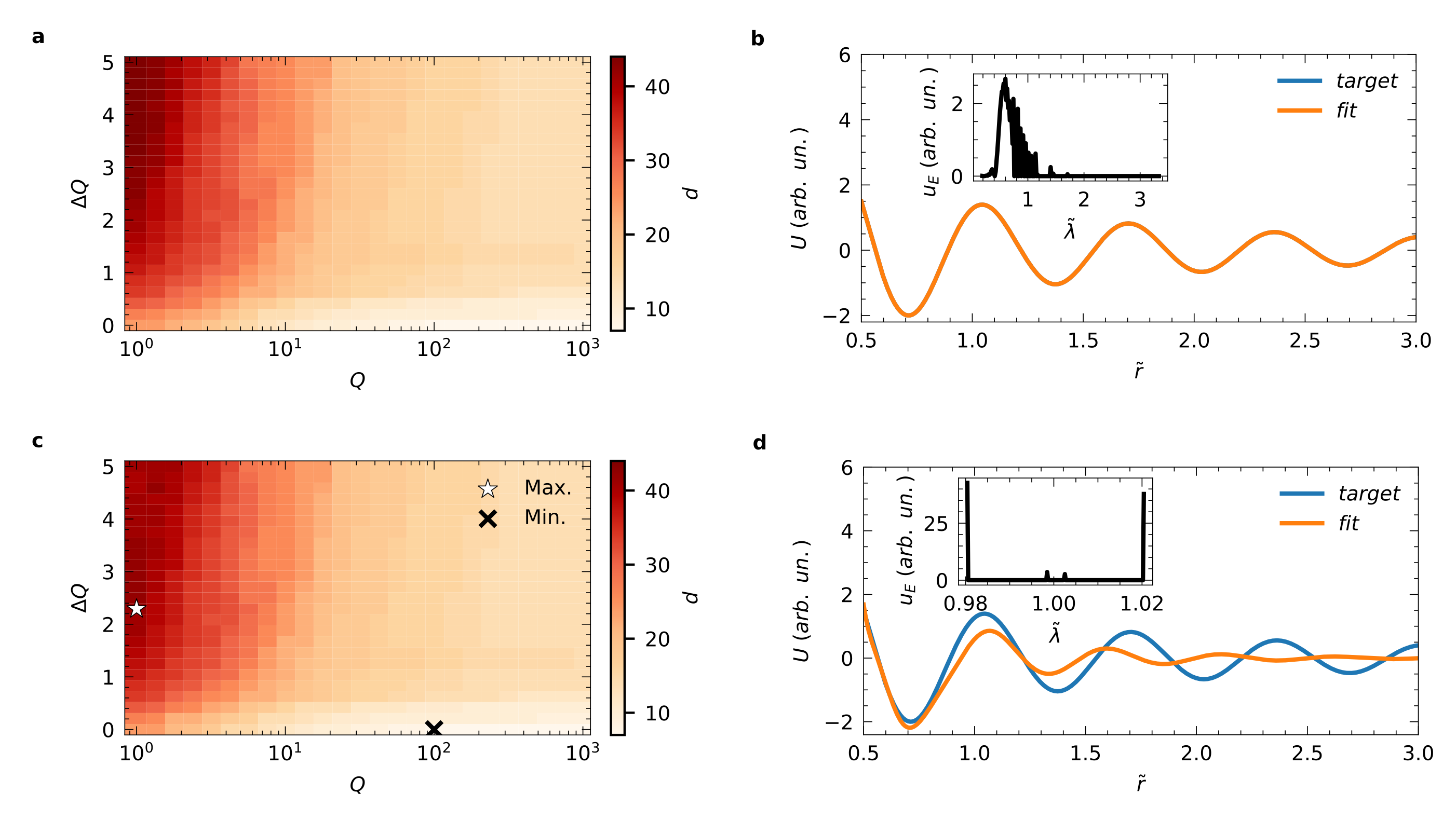}
    \caption{\textbf{a}, estimated dimension of the space of fittable function using the SVD. \textbf{c}, estimated dimension of the space of fittable functions with positive coefficients only. \textbf{b}, same example as in Figure \ref{fig:figure_SHU}\textbf{b}, with the particles polarizabilities giving the maximum positive dimension $d=42$ (star in \textbf{c}). \textbf{d}, same example but with the particle polarizabilities yielding the minimal positive dimension $d=7$ (cross in \textbf{c}).}
    \label{fig:dimension}
\end{figure*}

Figure \ref{fig:dimension}\textbf{a} shows the dimension of the fittable space as a function of the quality factor $Q$ and detuning parameter $\Delta Q$ describing the particles. It is computed with $N_l=200$ dimensionless wavelengths $\tilde\lambda_i\equiv\lambda_i/\lambda_0$, where $\lambda_0$ is the electric resonance wavelength, in the interval $\tilde{\lambda}_i \in \left[\left({1+\Delta+2/Q}\right)^{-1},\,{\max(1-2/Q,0.3)}^{-1}\right]$. The considered dimensionless distances $\tilde r=r/\lambda_0$ are discretized in $N_d=1000$ points in the interval $\tilde r \in \left [0.5, 3.0 \right ]$. The threshold is set to be $t=10^{-5}$. The dimension is high in the region where the quality factor $Q$ is between 1 and 10 and the detuning parameter multiplied by the quality factor $\Delta Q$ is between 2 and 5.

We remark that changing the threshold to $t=10^{-3}$ lowers the maximum dimensionality from $d=42$ to $d=23$, indicating a possible logarithmic dependence of $d$ on the arbitrary threshold $t$.

However, the dimension calculation presented above considers all possible linear combinations, without restricting to those with positive energy densities. To address the effect of this constrain, one can compute the SVD of the matrix $A$ using only the potentials $U_i$ for which $-U_i$ can be reasonably well fitted with the potentials at other wavelengths. Specifically, the absolute maximum fitting error over all positions $r_j$ must not exceed 10\%. Figure \ref{fig:dimension}\textbf{c} shows the same map as Figure \ref{fig:dimension}\textbf{a}, but with the corrected method to consider only positive spectral energy density components, showing a similar behavior but with a maximum at $d=42$, slightly lower than the one shown on Figure \ref{fig:dimension}\textbf{a}.
Interestingly, the regions of high effective dimensionality occur for small values of $Q$ and $\Delta\cdot Q$ that correspond to electric and magnetic resonances in close proximity or overlapping. This suggests that it is the interference between electric and magnetic resonance that provides the necessary degrees of freedom to expand the possible classes of interactions.

In order to illustrate this behavior, Figure \ref{fig:dimension}\textbf{b,d} shows the potential shown on figure \ref{fig:figure_SHU}\textbf{a}, but with the particles polarizabilities placed at the point of the maximal dimension ($d=42$, \ref{fig:dimension}\textbf{b})  and the minimal one ($d=7$, \ref{fig:dimension}\textbf{d}). In the former case, the fit reproduces very well the shape of the target potential whereas, in the latter, the shape is not well retrieved for distances bigger than $\tilde r\equiv r/ \lambda_0=0.75$.

\section{Discussion}
We demonstrated a method to tune colloidal pair interactions using artificial random light fields by optimizing the spectral energy density. This approach allows precise control over light-induced interaction potentials. 

The proposed fitting procedure effectively selects the optimal spectral energy density, achieving low error rates for various interaction potentials as for instance the electrostatic double layer potential, the Lennard-Jones potential or the potential allowing to get Stealthy Hyperuniform point patterns. Moreover, an additional study described in Appendix A shows that even if a small Gaussian noise is added or if a single-wavelength contribution is removed from the spectral energy density, the fitting procedure is quite robust, keeping at least the qualitative behavior of the fitted potential.

The dimension analysis of the fittable space of functions, indicates a wide range of attainable interaction potentials. The dimension is maximum for low quality factors and is not much affected by the restriction to only positive coefficients. This suggests that the interference between electric and magnetic dipoles plays an important role in the ability to generate pre-designed pair interactions.

\section*{Acknowledgements}
We acknowledge useful and stimulating discussions with Manuel Marqués, Bart van Tiggelen and Nicolas Cherroret. Authors acknowledge the financial support from Schweizerischer Nationalfonds zur Förderung der Wissenschaftlichen Forschung (197146).

\appendix
\section{Robustness of the tuning procedure}
\subsection{Absence of some spectral energy density components}
In this section, we study the effect of removing some contributing frequencies to the random field. For this, we consider the examples given in Figures \ref{fig:fig_DL}\textbf{b} and \ref{fig:figure_SHU}\textbf{b} (showing fits with $U^{DL}$ and $U^{SHU}$). For each of these two examples, we sorted the spectral energy density components by their magnitude and removed the first, second, or third smallest of them before computing the pair interaction using eq. \eqref{discrete_pot}. We then compared the obtained results with the target potential by computing $|U(r)-U_t(r)|$.
\begin{figure}
    \centering
    \includegraphics[width=1\columnwidth]{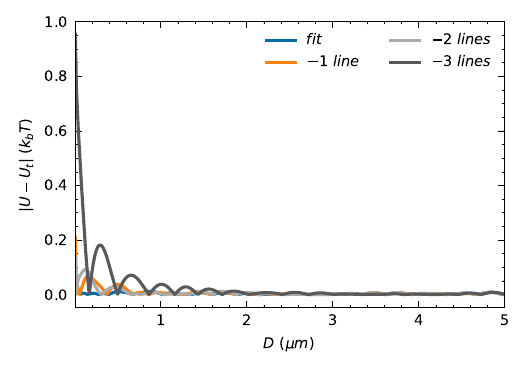}
    \caption{Error $|U(r)-U^{DL}(r)|$ as a function of $D$ for the complete fitting procedure and the same procedure where the first, second and third smallest contribution to the spectral energy density are removed.}
    \label{fig:rem_DL}
\end{figure}
Figure \ref{fig:rem_DL} shows the results obtained by setting the target potential to $U_t=U^{DL}$. An increase in the amplitude of the error with the number of removed spectral energy density contributions can be noticed. However, when removing one or two lines, this error stays close to the error of the complete fitting procedure and way lower than $k_BT$. Similarly, Figure \ref{fig:rem_SHU} is showing that the fitting procedure with $U_t=U^{SHU}$ is quite robust when the first or second smallest contributions to the random field are removed.
\begin{figure}
    \centering
    \includegraphics[width=1\columnwidth]{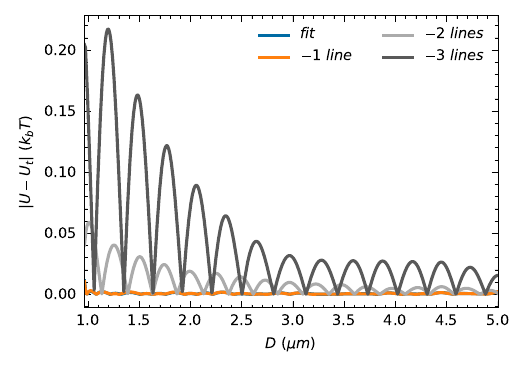}
    \caption{Error $|U(r)-U^{SHU}(r)|$ as a function of $D$ for the complete fitting procedure and the same procedure where the first, second and third smallest contribution to the spectral energy density are removed.}
    \label{fig:rem_SHU}
\end{figure}

\subsection{Introduction of random Gaussian noise to the spectral energy density}
Similarly, we investigate the effect of introducing random Gaussian noise to the spectral energy density. Specifically, we consider the examples given in Figures 2\textbf{b} and 4\textbf{b}, which show fits with $U^{DL}$ and $U^{SHU}$, respectively. For each of these examples, we add Gaussian-distributed noise with zero mean and varying standard deviations. Figures 8 and 9 illustrate the obtained results for $U^{DL}$ and $U^{SHU}$, respectively. The standard deviation of the noise is set to 1\%, 3\%, 5\%, and 10\% of the maximum of the spectral energy density. The plain lines correspond to the results of the fitting procedure without noise, while the grey areas represent the range between the maximum and minimum differences for each distance $ D $ across 10,000 realizations. The results show that introducing small random Gaussian noise (up to 3\%) keeps the error relatively small and preserves the general behavior of the potential, ensuring reliable control over the induced pair interactions.

\begin{figure}
    \centering
    \includegraphics[width=1\columnwidth]{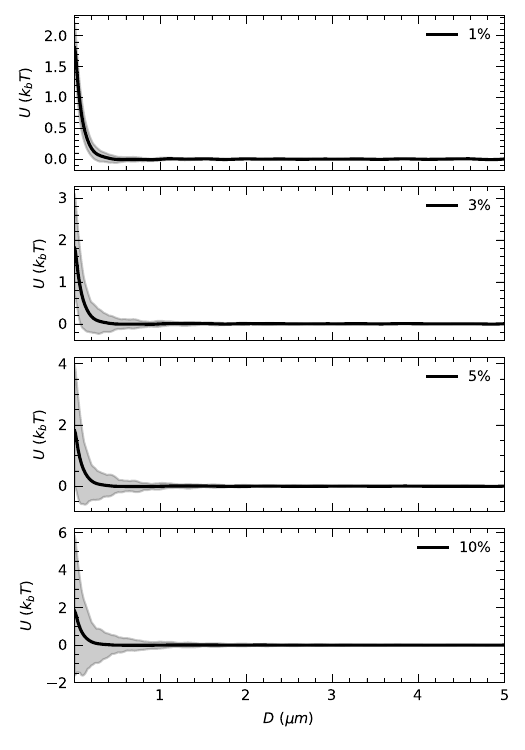}
    \caption{Error due to the introduction of random Gaussian noise on the spectral energy density, taking $U_t=U^{DL}$. From top to bottom, the standard deviation of the random noise is set to be $1\%, 3\%, 5\%$ and $10\%$ of the maximum of the spectral energy density. Plain line correspond to the results of the fitting procedure as presented in Figure \ref{fig:fig_DL}\textbf{b} and the grey area corresponds to the area between the maximum and minimum difference for each $D$ on a sample of $10000$ realizations.}
    \label{fig:gauss_DL}
\end{figure}
\begin{figure}
    \centering
    \includegraphics[width=1\columnwidth]{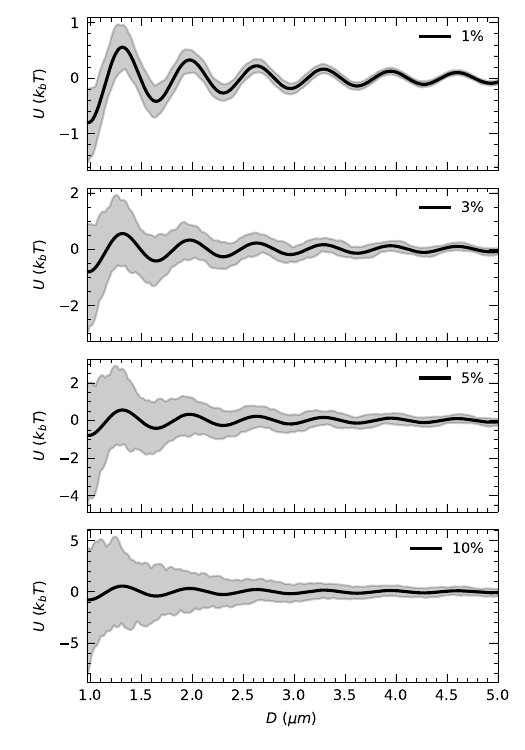}
    \caption{Error due to the introduction of random Gaussian noise on the spectral energy density, taking $U_t=U^{SHU}$. From top to bottom, the standard deviation of the random noise is set to be $1\%, 3\%, 5\%$ and $10\%$ of the maximum of the spectral energy density. Plain line correspond to the results of the fitting procedure as presented in Figure \ref{fig:figure_SHU}\textbf{b} and the grey area corresponds to the area between the maximum and minimum difference for each $D$ on a sample of $10000$ realizations.}
    \label{fig:gauss_SHU}
\end{figure}
\section{Parametrization of the electric and magnetic polarizabilities}
In order to compute the dimension of the fittable subspace in a general way, it is necessary to establish a general description of the particle electric and magnetic dipole response (notice that it is described in more detail in \cite{muster2025pure}). To this end, we model the electric or magnetic polarizability $\alpha_d$, $d=e,m$, by a Lorentzian lineshape

\begin{equation}
    \alpha_d\left(\omega\right) = \frac{6 \pi}{k^3} \frac{\gamma \omega}{\omega_0^2 - \omega^2 - i \omega \gamma},
\end{equation}
where $\omega_0$ is the resonance frequency and $\gamma$ is its damping rate. Notice that this expression for the polarizability satisfies the optical theorem \cite{jones2015optical} $k\,\mathrm{Im}\{\alpha_d\} = {k^4} |\alpha_d|^2/{6\pi}$.

To reduce the parameter space required to describe the polarizability, frequency and wavelength can be expressed in terms of their values at electric resonance as
\begin{equation}
    \tilde{\omega}\equiv \frac{\omega}{\omega_0}, \quad \tilde{\lambda} \equiv \frac{\lambda}{\lambda_0} = \tilde{\omega}^{-1}.
\end{equation}

In addition, we define the resonance's quality factor as $ Q \equiv \frac{\omega_0}{\gamma}$, which let us rewrite the polarizability as
\begin{equation}
    \frac{\alpha_d(\omega)}{\lambda_0^3} = \frac{3}{4 \pi^2  Q} \frac{\tilde{\omega}^{-2}}{1 - \tilde{\omega}^2 - i \tilde{\omega}/Q}.
\end{equation}
With this scaling, a single resonance is specified by its quality factor $Q$. Since the particles considered in this work support both electric and magnetic dipoles polarizabilities, we introduce two quality factors, $Q_e$ and $Q_m$, along with a dimensionless detuning parameter that sets their relative spectral position defined as $\Delta = \frac{\omega_0^m - \omega_0^e}{\omega_0^e}$, so that ${\omega_0^m}/{\omega_0^e} = 1 + \Delta$. In this work, we will set $Q=Q_e=Q_m$ in order to reduce the dimension of the parameter space.

\bibliography{biblio.bib}

\end{document}